\PassOptionsToPackage{table,xcdraw}{xcolor}
\documentclass[10pt,twocolumn,letterpaper]{article}

\usepackage{iccv}              

\usepackage{times}
\usepackage{epsfig}
\usepackage{graphicx}
\usepackage{amsmath}
\usepackage{amssymb}
\usepackage{booktabs}
\usepackage{caption}
\usepackage{color}
\usepackage{multirow}
\usepackage{colortbl} 
\usepackage{enumitem}
\usepackage[ruled, lined,noend, linesnumbered, commentsnumbered]{algorithm2e}
\usepackage{tikz}
\usepackage{float}
\usepackage{fontawesome5}
\usepackage[accsupp]{axessibility}
\graphicspath{{figs/}}

\newcommand{\mysection}[1]{\vspace{2pt}\noindent\textbf{#1}}

\usepackage[normalem]{ulem}
\useunder{\uline}{\ul}{}
\usepackage{graphicx}

\usepackage{xspace}
\newcommand{\Wpl}{$\mathcal{W}^{+}$\xspace}

\newcommand{\Fk}{$\mathcal{F}_k$\xspace}

\definecolor{customgreen}{RGB}{73, 187, 121}

\definecolor{secLock}{HTML}{00b050}
\definecolor{unsecLock}{HTML}{ff0000}
\definecolor{commentsColor}{RGB}{219, 48, 122}

\SetCommentSty{mycommfont}
\let\oldnl\nl
\newcommand{\nonl}{\renewcommand{\nl}{\let\nl\oldnl}}

\usepackage[normalem]{ulem}
\useunder{\uline}{\ul}{}

\usepackage[precision=2, unit=mm]{lengthconvert}

\usetikzlibrary{patterns}

\definecolor{figblue}{HTML}{1560bd}
\definecolor{figred}{HTML}{a9203e}
\definecolor{figpurple}{HTML}{8844aa}
\definecolor{figgreen}{HTML}{00693e}

\definecolor{iccvblue}{rgb}{0.21,0.49,0.74}
\usepackage[pagebackref,breaklinks,colorlinks,allcolors=iccvblue]{hyperref}
\usepackage{subcaption}
\usepackage{booktabs} 

\def\paperID{1186} 
\def\confName{ICCV}
\def\confYear{2025}

\title{
    \raisebox{-0.3\height}{\includegraphics[width=1cm]{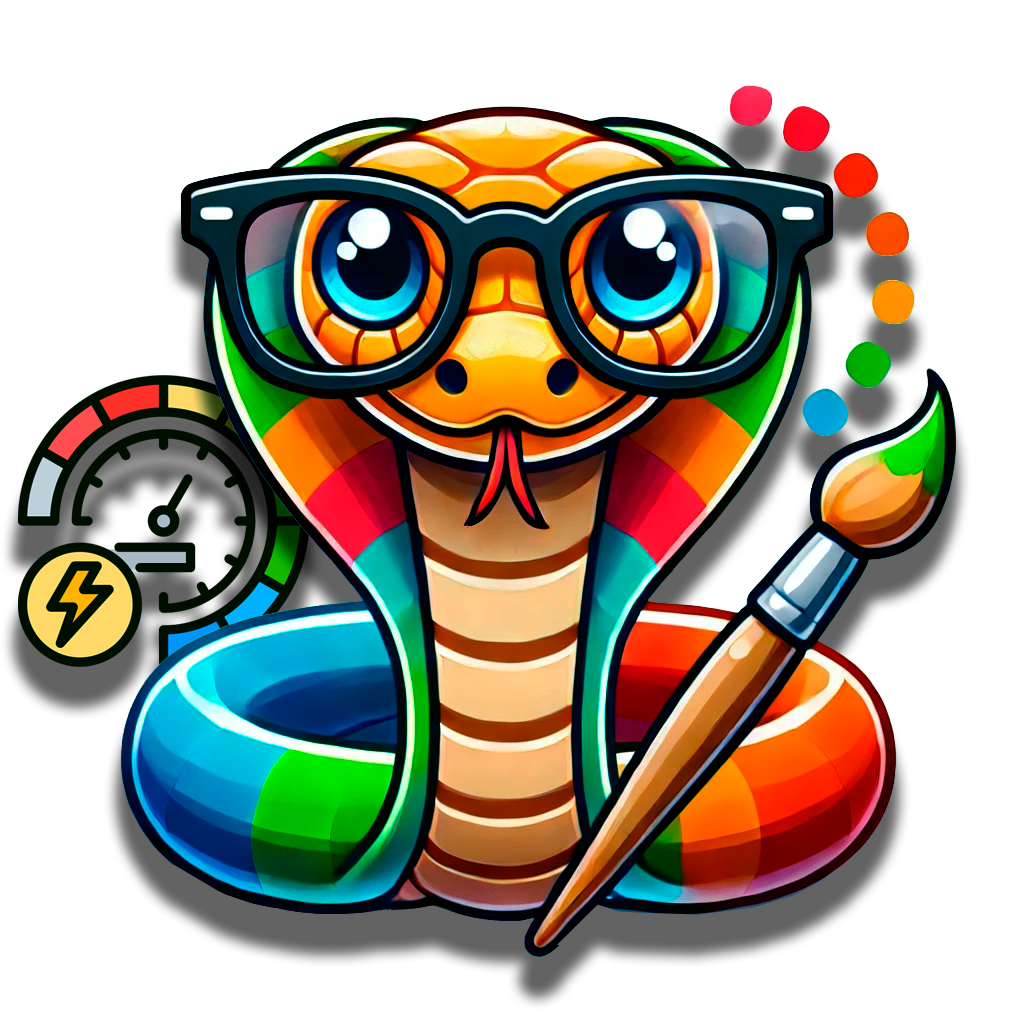}} 
    MambaStyle: Efficient StyleGAN Inversion for Real Image Editing with State-Space Models
}

\author{Jhon Lopez$^{1,2, \dagger}$, Carlos Hinojosa$^{2,*}$, Henry Arguello$^{1}$, Bernard Ghanem$^{2}$\\
$^{1}$Universidad Industrial de Santander; \ \ $^{2}$KAUST \\
}

\begin{document}
\maketitle

\def\thefootnote{*}\footnotetext{Project lead; $\dagger$ Work done during internship at KAUST.}

\begin{abstract}
The task of inverting real images into StyleGAN's latent space to manipulate their attributes has been extensively studied. However, existing GAN inversion methods struggle to balance high reconstruction quality, effective editability, and computational efficiency. In this paper, we introduce \textbf{MambaStyle}, an efficient single-stage encoder-based approach for GAN inversion and editing that leverages vision state-space models (VSSMs) to address these challenges. Specifically, our approach integrates VSSMs within the proposed architecture, enabling high-quality image inversion and flexible editing with significantly fewer parameters and reduced computational complexity compared to state-of-the-art methods. Extensive experiments show that MambaStyle achieves a superior balance among inversion accuracy, editing quality, and computational efficiency. Notably, our method achieves superior inversion and editing results with reduced model complexity and faster inference, making it suitable for real-time applications.
\end{abstract}
\vspace{-3 mm}
    
\section{Introduction}
\label{sec:intro}

Recent progress in Generative Adversarial Networks (GANs) \cite{goodfellow2014generative} has demonstrated impressive success in generating high-quality synthetic images. Among recent models, StyleGAN \cite{karras2019style,karras2020analyzing, karras2021alias} stands out due to its ability to generate photorealistic images and provide a semantically rich latent space, which enables controllable image editing \cite{abdal2020image2stylegan++}. Specifically, once an image is represented in StyleGAN's latent space, we can control different semantic attributes by modifying its latent representation. The task of finding such latent representation for a \textit{real} image is known as GAN inversion \cite{zhu2016generative, xia2022gan}.



\begin{figure}
    \centering
    \includegraphics[width=\columnwidth]{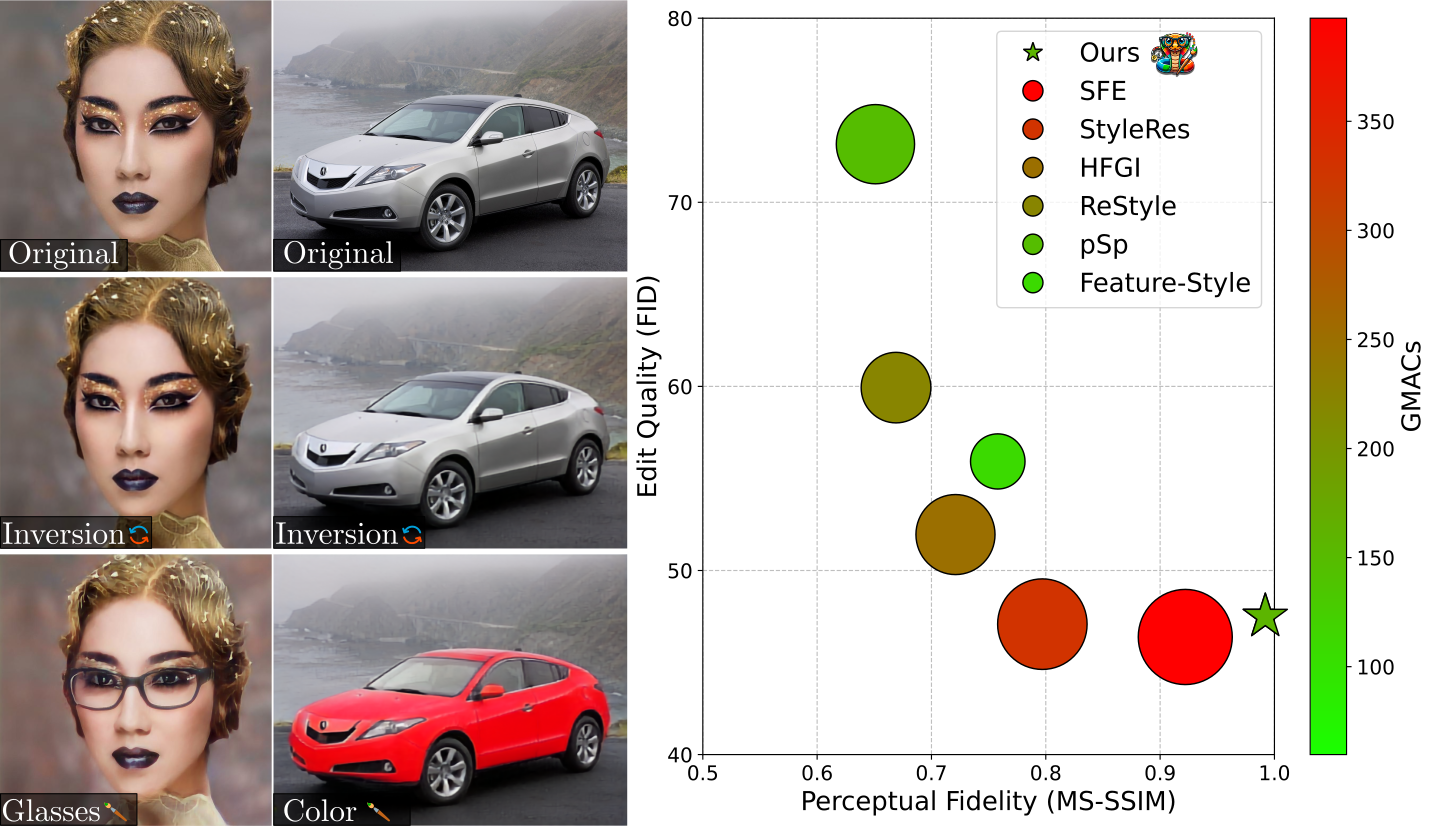}
    \caption{\small Our method encodes real images into the StyleGAN latent space, applies edits and synthesizes the edited images. We compare our MambaStyle with prior methods in the right plot in terms of MS-SSIM$\downarrow$ for inversion quality and FID$\downarrow$ for editing capability. Larger markers indicate a higher model parameter count.}
    \label{fig:intro_fig}
    \vspace{-6 mm}
\end{figure}

In general, GAN inversion methods can be categorized into optimization-based and encoder-based approaches. Optimization-based methods iteratively adjust the latent code to match a target image, often achieving high inversion quality \cite{abdal2019image2stylegan,abdal2020image2stylegan++}. However, they have two main drawbacks: (1) they produce overfitted latent codes that may deviate from the original latent space distribution, leading to poor editability; and (2) since they require a separate optimization process for each image, their high computational cost makes them impractical for real-time applications. Encoder-based methods, on the other hand, aim to directly map real images to the GAN's latent space through a neural network, allowing for faster inference \cite{pehlivan2023styleres,bobkov2024devil,hu2022style,wei2022e2style}. Although encoder-based methods improve efficiency, they often struggle to match the image inversion accuracy of optimization-based methods and to achieve a good balance between inversion quality and editability, commonly referred to as the distortion-editability trade-off \cite{tov2021designing}. Recent approaches have implemented more complex encoder architectures that map images into the feature space of StyleGAN, denoted as $\mathcal{F}_k$. While these methods achieve a better distortion-editability trade-off, they typically increase model complexity and parameters (see Fig. ~\ref{fig:intro_fig} - right) \cite{bobkov2024devil}. Thus, the need for a more efficient architecture persists.

Recently, Mamba \cite{gu2023mamba, dao2024transformers}, a new architecture derived from state-space models (SSMs) \cite{gu2020hippo,gu2021efficiently}, has emerged as a computationally and memory-efficient backbone for constructing effective deep network architectures across various tasks, including language processing \cite{wang2024mambabyte} and vision tasks such as image classification and image restoration \cite{shi2024vmambair}, among others. This development hints at a potential solution for designing an encoder-based GAN inversion method that achieves high-quality inversion, effective editability, and efficiency. Yet, despite its success across vision applications, no prior work has applied SSM-based modules within encoder networks for GAN inversion.




\mysection{Paper Contributions.} Inspired by the success of SSMs, we propose \textbf{MambaStyle}, a single-stage encoder-based GAN inversion and editing framework that leverages SSM to achieve high reconstruction quality and effective editability, with significantly fewer parameters, reduced computational complexity, and faster inference compared to current approaches. Specifically, our main contributions are:

\begin{enumerate}
    \item We propose \textbf{MambaStyle}, a novel encoder-based GAN inversion and editing framework that leverages vision state-space models for efficient latent representation learning, significantly reducing parameters, inference time, and computational complexity compared to existing encoder-based methods, providing a more efficient and faster alternative.
    \item Our proposed method employs a single-stage training approach that leverages paired data samples generated from noise using pretrained StyleGAN mapping and generator networks. This method eliminates the need for multi-phase training and the use of additional pretrained encoders, such as e4e.
    \item Through extensive experiments, we demonstrate that our approach achieves an optimal balance between high-quality inversion, editing performance, and model efficiency, outperforming current state-of-the-art methods across multiple metrics.
\end{enumerate}

Our proposed MambaStyle framework consists of two key components: a multi-scale \textbf{Encoder} for feature extraction and a novel \textbf{Fuser} module. The Fuser directly integrates editing information into the spatial feature maps produced by the encoder, enhancing fine-grained details while preserving global structural coherence during inversion and editing. Section ~\ref{sec:related_works} introduces the related works; in Section ~\ref{sec:method}, we describe our proposed framework in detail; and Section ~\ref{sec:experiments} comprehensively validates our approach through different experiments and ablation studies.
\section{Related Work}
\label{sec:related_works}

The goal of GAN inversion is to find a latent code for a real image by projecting it into the GAN's embedding space such that the GAN can reconstruct the original image from this latent representation \cite{zhu2016generative}. This latent code can then be manipulated to change the semantic attributes of the image \cite{abdal2020image2stylegan++, shen2020interpreting}. Specifically, such manipulation involves moving along specific directions within the GAN's latent space to apply targeted edits, such as adjusting facial expressions, modifying age, or adding objects, while maintaining the core identity of the original image. However, effective manipulation relies on high-quality inversion, since a well-inverted latent code ensures that edits retain the image's structural and identity consistency. With the development of StyleGAN models \cite{karras2019style, karras2020analyzing, karras2021alias}, GAN inversion has gained popularity due to StyleGAN's semantically rich and disentangled features, which enable controlled and realistic edits \cite{wu2021stylespace}. This has led to the development of methods focused on achieving both high reconstruction fidelity and editing quality, broadly categorized into optimization-based \cite{bhattarai2024triplanenet,roich2022pivotal, abdal2020image2stylegan++} and encoder-based GAN inversion methods \cite{bobkov2024devil,pehlivan2023styleres,liu2023delving,yao2022feature,wang2022high,alaluf2022hyperstyle,lopez2024privacy,dinh2022hyperinverter,hu2022style,bai2022high,alaluf2021restyle,richardson2021encoding,tov2021designing}.

\mysection{Optimized-based Inversion Methods.} These methods iteratively adjust the latent code to closely match a target image by minimizing a reconstruction loss. This approach often achieves high-quality inversion results, as the latent code is specifically optimized for each image, capturing fine details and features \cite{abdal2019image2stylegan,abdal2020image2stylegan++,zhu2020improved, roich2022pivotal, bhattarai2024triplanenet}. For instance, in \cite{roich2022pivotal}, the authors propose not only to optimize the latent code but also to fine-tune the generator parameters. This approach helps prevent the latent code deviating from the StyleGAN manifold during optimization, which could affect the editability of the generated image. Although highly effective, these methods can be computationally slow, since each image requires a separate optimization process, making them impractical for real-time applications.


\begin{figure*}[t!]
\vspace{-3 mm}
   \includegraphics[width=\textwidth]{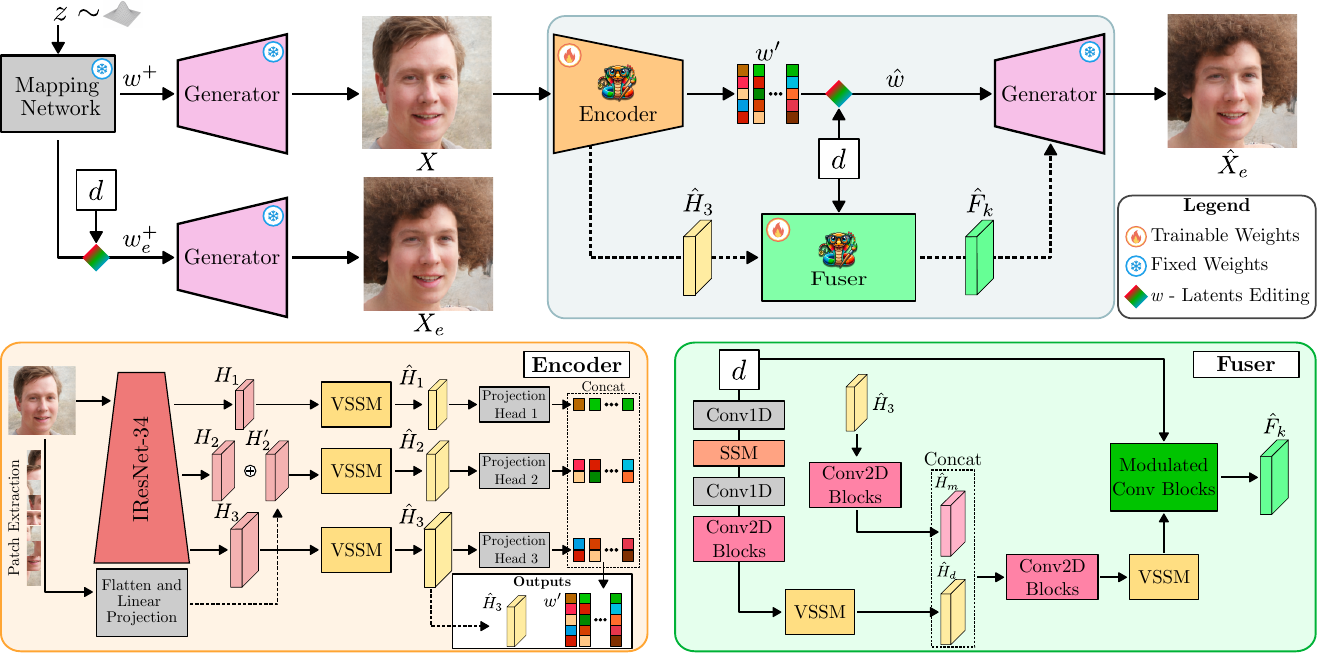}
  \caption{During \textbf{training}, we leverage the pretrained StyleGAN2 Generator and Mapping network to generate an image $X$ and its edited version $X_e$  from a random noise vector $z$ and an editing direction $d$. Then, we learn the latent vectors $\hat{w} \in \mathcal{W}^{+}$ and features $\hat{F}_k$ $\in \mathcal{F}_{k}$ using our proposed \textit{MambaStyle} architecture. This architecture comprises (i) a Multi-scale Mamba-based \textit{Encoder} that encodes the input image $X$ and produces $w'$ and features $\hat{H}_3$; and (ii) a \textit{Fuser} module, which integrates the editing direction $d$ with features $\hat{H}_3$ to generate $\hat{F}_k$. Using $\hat{w}$ and $\hat{F}_k$, the pretrained StyleGAN2 model synthesizes an image with $G(\hat{F}_k, \hat{w})$. Our proposed framework is flexible, allowing both image inversion and editing conditioned on the $d$  direction. Specifically, setting $d = \mathbf{0}$ enables image inversion, reconstructing an approximation of the original image $\hat{X}$; when $d \neq \mathbf{0}$, our framework performs image editing, generating the edited image $\hat{X}_e$ based on the specified direction $d$. During \textbf{inference}, we follow the same pipeline, with the encoder taking the real image $X$ from the target dataset.}
  \label{fig:proposed}
  \vspace{-5 mm}
\end{figure*}

\mysection{Encoder-based Inversion Methods.} These methods aim to directly map real images to the GAN's latent space using a trained neural network. This allows for faster inference since it eliminates the need for iterative optimization on each image. However, they struggle to achieve both high-quality reconstruction and precise image edits. Early approaches focused on mapping to simpler spaces, which allowed for good editability but sacrificed inversion accuracy \cite{pidhorskyi2020adversarial, richardson2021encoding, tov2021designing, zhu2020domain}. 
Recent methods train an encoder to map images to StyleGAN's high-dimensional feature space, $\mathcal{F}_k$ \cite{bobkov2024devil, pehlivan2023styleres, wang2022high, yao2022feature}, achieving improved editability and reconstruction quality among encoder-based approaches. For instance, the StyleFeatureEditor (SFE) \cite{bobkov2024devil} method employs a two-stage training process where an ``Inverter'' is first trained to extract reconstruction features from the input image, followed by training a ``Feature Editor'' to transform these features based on the desired edit. Although these approaches improve both inversion and editing quality, they increase model parameters and complexity. 

\mysection{State-Space Models.} State-space models (SSMs) are a mathematical representation of dynamic systems that model the input-output relationship through a hidden state. Recently, SSMs have emerged as an alternative network architecture to model long-range dependency \cite{gu2020hippo,gu2021efficiently}. Compared to Transformer-based networks \cite{dosovitskiy2021an,wolf2020transformers}, which have quadratic complexity relative to sequence length, SSMs offer significantly greater computational and memory efficiency. Among recent advances in SSMs, Gu et al. \cite{gu2021efficiently} introduced structured state-space models (S4) to improve computational efficiency by representing the state matrix as a sum of low-rank and normal matrices. Building on S4, Mamba \cite{gu2023mamba} introduces the core operation S6, an input-dependent selection mechanism for S4, which achieves linear scaling with sequence length. Recently, the authors in \cite{liu2024vmamba} adapted the S6 algorithm for visual data, enabling Mamba-based vision state-space models (VSSM) to be applied to various vision tasks, including image classification \cite{liu2024vmamba,zhu2024vision}, image restoration \cite{shi2024vmambair}, and segmentation \cite{xing2024segmamba}. However, to the best of our knowledge, no prior work has explored the use of VSSM for GAN inversion and editing.
\vspace{-2 mm}

\section{Proposed Method}
\label{sec:method}






In StyleGAN-based inversion and editing methods, two primary latent spaces are commonly used: the \Wpl space and the \Fk feature space, which combines \Wpl with the feature outputs of the $k$-th convolutional layer in StyleGAN. The \Wpl space, which comprises the generator's layer-specific latent vectors, is well-suited for making precise edits due to its disentangled representation. However, reconstructions using \Wpl suffer from lower fidelity, often losing fine details of the original image. In contrast, encoding the image into higher feature maps, such as the \Fk space, enables accurate reconstruction, preserving fine details. Unfortunately, this high-fidelity representation in \Fk is difficult to edit, limiting its flexibility for editions. To address these limitations and improve efficiency, we propose a single-stage approach that combines both inversion and editing. 
\vspace{-3 mm}


\mysection{Overview.} Figure ~\ref{fig:proposed} presents an overview of our proposed framework. In our approach, we generate an original image $X$ and its edited version $X_e$  by leveraging the capabilities of StyleGAN2's pretrained mapping network and generator $G$. First, we sample a random noise vector $z$ from a Gaussian distribution, then fed into StyleGAN2's mapping network to produce a latent code in the intermediate \Wpl space, represented as  $w^{+}$. This latent code  $w^{+}$  is input to the generator, resulting in the original image  $X = G(w^{+})$. To obtain an edited version of this image, we randomly sample an editing direction $d \in \mathcal{D}$ and apply it in the \Wpl space. Specifically, the latent code $w^{+}$ is adjusted by adding $d$ to produce a modified latent representation $w^+_e = w^{+} + d$. This edited latent code $w^+_e$ is then passed through the generator, yielding the edited image $X_e = G(w^+_e)$. By manipulating the latent vector in \Wpl space with a targeted direction $d$, our approach enables flexible, controllable editing that preserves underlying characteristics of the original image while modifying specific attributes according to the direction (see our supplementary for a comparison when using e4e \cite{tov2021designing} to obtain $w^{+}$). Given $X$, we efficiently learn the latent vectors $\hat{w} \in \mathcal{W}^{+}$ and features $\hat{F}_k$  using our proposed \textit{MambaStyle} architecture, which consists of two main components: (i) a Multi-scale Mamba-based \textit{Encoder} that jointly encodes relevant information into $\mathcal{W}^{+}$ and $\mathcal{F}$ spaces, producing outputs $\hat{w}$ and $\hat{H}_3$, respectively, and (ii) a Mamba-based module called the \textit{Fuser}, which combines the editing direction $d$ with the feature maps $\hat{H}_3$ extracted from the Encoder to generate $\hat{F}_k$. Finally, we obtain the inverted/edited image with $G(\hat{F}_k,\hat{w})$, where $\hat{F}_k$  serves as a skip connection (added) to StyleGAN at layer $k=9$.


\subsection{Preliminaries}
\label{sec:preliminaries}
State-Space Models (SSMs), such as S4 \cite{gu2021efficiently}, learn to map a 1-D sequence $x(t) \in \mathbb{R}$ to another 1-D sequence $ y(t) \in \mathbb{R}$ as output, through an internal state-space $h(t) \in \mathbb{R}^N$, using a linear differential equation:
{\small
\begin{equation}
\setlength{\belowdisplayskip}{4pt}
\setlength{\abovedisplayskip}{4pt}
h'(t) = Ah(t) + Bx(t), \quad y(t) = Ch(t) + Dx(t),
\label{eq:cont_ssm}
\end{equation}}

\noindent where the matrices $A \in \mathbb{R}^{N \times N}, B \in \mathbb{R}^{N \times 1}, C \in \mathbb{R}^{1 \times N}$, and $D \in \mathbb{R}$ are learnable. Since a continuous-time system is not suitable for digital computers and real-world data, we apply a discretization procedure to the system in Equation \ref{eq:cont_ssm}. Using the zero-order hold (ZOH) method, with  $\Delta$ as the discrete-time step and $\bar{A}$ and $\bar{B}$ as the discretized matrices, the system is transformed into the discrete-time form:
{\small
\begin{equation}
\setlength{\belowdisplayskip}{4pt}
\setlength{\abovedisplayskip}{4pt}
h_t = \bar{A} h_{t-1} + \bar{B} x_t, \quad y_t = C h_t + D x_t.
\label{eq:ssm}
\end{equation}}
A key limitation of SSMs lies in how well the state can compress the context information. To overcome this, Mamba \cite{gu2023mamba, dao2024transformers} proposes S6, which introduces an input-dependent selection mechanism to allow the system to select relevant information based on the input sequence. This is achieved by making $B$, $C$, and $\Delta$ functions of the input $x_t$, as:
{\small
\begin{equation}
\setlength{\belowdisplayskip}{4pt}
\setlength{\abovedisplayskip}{4pt}
B = \text{Lin}_B(x), \quad C = \text{Lin}_C(x), \quad \Delta = \text{Lin}_{\Delta}(x),
\end{equation}}
with Lin$_{*}$ being a linear, fully-connected layer. 


\subsection{MambaStyle Architecture}
Our proposed architecture consists of two main modules: Multi-scale Mamba-based Encoder and Fuser. 

\begin{figure}
    \centering
    \includegraphics[width=\columnwidth]{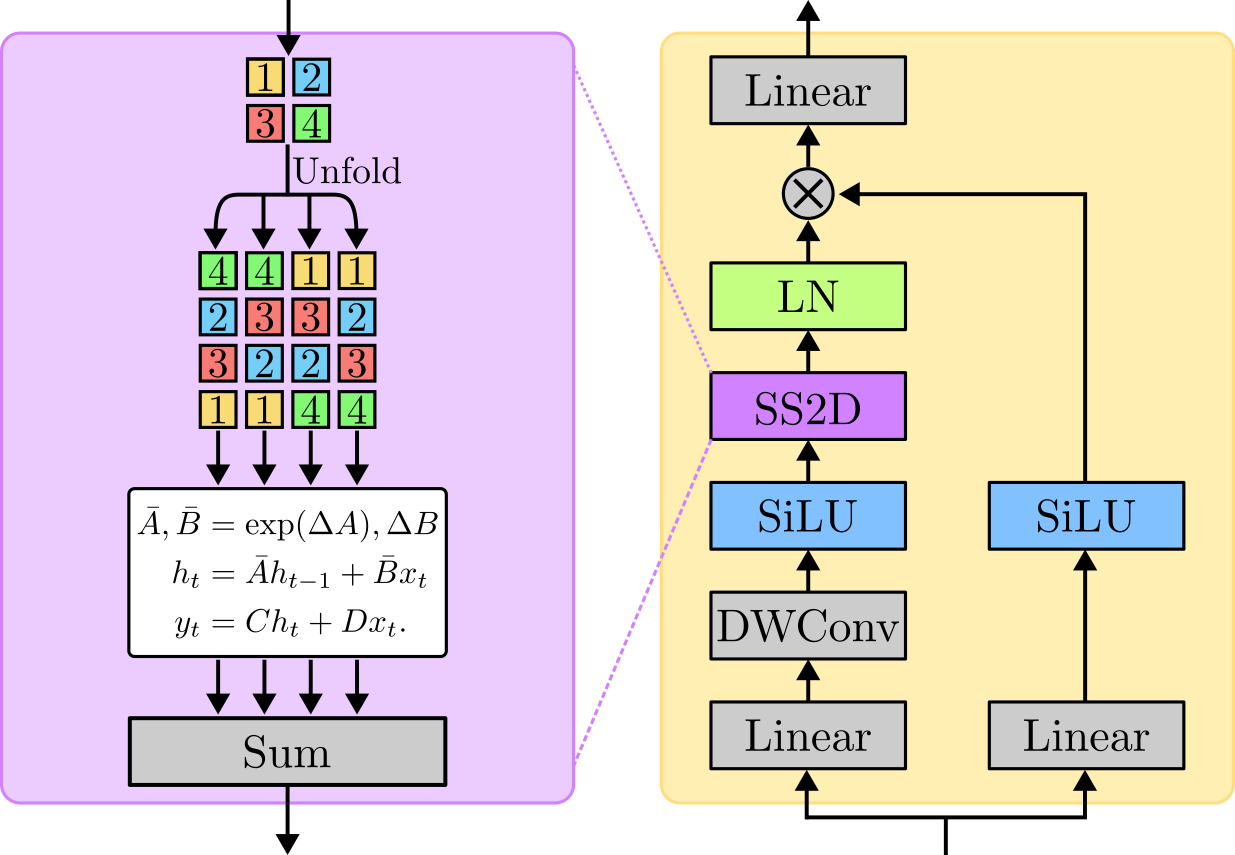}
    \caption{(Right) Vision State-Space Module (VSSM), and (Left) the 2D Selective Scan Submodule (SS2D).}
    \vspace{-15pt}
    \label{fig:mamba_block}
\end{figure}

\mysection{Encoder.} Our proposed encoder $E$ consists of an IResNet-34 backbone, three Vision SSM (VSSM) modules, and three projection heads, each comprising batch normalization, 3D average pooling, and linear fully connected layers. Since the S6 algorithm introduced in Section \ref{sec:preliminaries} operates on 1D language sequences and is not directly applicable to 2D images, we adopt the 2D-selective-scan (SS2D) operation from \cite{liu2024vmamba} to adapt S6 for visual data, which serves as the basis of the VSSM module. Specifically, the SS2D method \cite{liu2024vmamba} expands image patches in four directions, generating four unique feature sequences. Each feature sequence is then processed by S6, where the state and output are calculated according to Eq. \ref{eq:ssm}. Finally, the processed features are combined to construct the comprehensive 2D feature map. The adopted VSSM is illustrated in Fig. \ref{fig:mamba_block}.

Given the input image $X$, the IResNet-34 backbone extracts three hierarchical feature maps, denoted as $H_1$, $H_2$, and $H_3$, which represent multi-scale features at different layers of the network. Simultaneously, $X$ is divided into fixed-size patches, each of which is flattened and linearly projected to obtain embeddings. Positional embeddings are then added to retain spatial information, resulting in the feature map $H'_2$, which is added to $H_2$. Each of these feature maps is subsequently processed by the VSSM modules, producing the refined feature maps $\hat{H}_1$, $\hat{H}_2$, and $\hat{H}_3$, that capture multi-scale and contextual information, enhancing the feature representations. Finally, these refined features are fed into projection heads, and their outputs are concatenated to form the latent vector representation $w' \in \mathcal{W}^{+}$. In summary, our encoder $E$ takes $X$ as input and outputs the latent representation $w'$ and feature $\hat{H}_3$:
{\small
\begin{equation}
\setlength{\belowdisplayskip}{4pt}
\setlength{\abovedisplayskip}{4pt}
    (w', \hat{H}_3)=E(X).
    \label{eq:encoder}
\end{equation}}


\mysection{Fuser.} Our proposed Fuser module $U$ is designed to preserve fine details related to the person's identity and background while enabling the fusion of reconstruction features $\hat{H}_3$ extracted from the Encoder $E$ with the latent space direction $d$. 
Given the input latent space direction $d$, we extract and transform features using a sequence of layers, including 1D convolutions, a 1D SSM module, 2D convolutions, and a VSSM module, to adapt it for fusion with $\hat{H}_3$, resulting in the feature maps $\hat{H}_d$. The SSM module here uses the same architecture as the VSSM shown in Fig. \ref{fig:mamba_block}, but it employs the original S6 algorithm rather than SS2D. Simultaneously, $\hat{H}_3$ is processed by 2D convolutional blocks to create an intermediate feature $\hat{H}_m$. Then, $\hat{H}_d$ and $\hat{H}_m$ are then concatenated, allowing the fusion of identity-preserving information with editing instructions. The concatenated feature is further refined through additional Conv2D blocks and a VSSM module. Finally, the fused features pass through Modulated Convolution Blocks \cite{karras2020analyzing}, which dynamically adjust the features based on the input direction $d$, producing the output $\hat{F}_k \in \mathcal{F}$. In summary, our proposed Fuser $U$ takes $\hat{H}_3$ and $d$ as input and outputs the features $\hat{F}_k$:
{\small
\begin{equation}
\setlength{\belowdisplayskip}{4pt}
\setlength{\abovedisplayskip}{4pt}
    \hat{F}_k = U(\hat{H}_3,d).
    \label{eq:fuser}
\end{equation}}



\mysection{Image Synthesis.} Our proposed framework enables both image inversion and editing, conditioned on the latent space direction $d$. Specifically, when $d = \mathbf{0}$ (a zero-valued tensor), the framework performs \textbf{image inversion}, reconstructing an approximation of the original image $\hat{X}$. Conversely, when $d \neq \mathbf{0}$, the framework performs \textbf{image editing}, synthesizing the image $\hat{X}_e$ according to the specified direction $d$. Formally, we synthesize an image as follows:
{\small
\begin{equation}
\setlength{\belowdisplayskip}{4pt}
\setlength{\abovedisplayskip}{4pt}
 G(\hat{F}_k, \hat{w})= 
    \begin{cases}
        \hat{X},& \text{if } d = \mathbf{0}\\
        \hat{X}_e,              & \text{otherwise}
    \end{cases},
\end{equation}}
\noindent where $G$ represents the pretrained StyleGAN2 generator, $\hat{F}_k$ are the features obtained from Eq. \ref{eq:fuser}, and $\hat{w} = w' + d$, where  $w'$  is obtained from Eq. \ref{eq:encoder}.





\subsection{Training Objectives}
Following prior works \cite{wei2022e2style, pehlivan2023styleres, bobkov2024devil, wang2022high}, we optimize our model using a weighted combination of loss functions to achieve high-fidelity image inversion and controlled editing. The reconstruction loss $\mathcal{L}_{\text{rec}}$ and perceptual loss $\mathcal{L}_{\text{perc}}$ \cite{zhang2018unreasonable} ensure accurate image recovery, while identity preservation $\mathcal{L}_{\text{id}}$ employing pretrained ArcFace model \cite{deng2019arcface}  and structure consistency $\mathcal{L}_{\text{struct}}$ leveraging a  pretrained U-Net model \cite{lee2020maskgan}, maintain subject identity and spatial coherence. Additionally, we introduce an editing loss $\mathcal{L}_{\text{e}}$ to enforce structured transformations in the latent space:  
{\small
\begin{equation}
\mathcal{L}_{\text{e}} = \| (\hat{w}_e - \hat{w}) - d \|_1,
\end{equation}}
where $d$ represents the predefined editing direction. The total loss function is given by:  
{\small
\begin{equation}
\mathcal{L}_{\text{total}} = \lambda_{1} \mathcal{L}_{\text{rec}} + \lambda_{2} \mathcal{L}_{\text{perc}} + \lambda_{3} \mathcal{L}_{\text{id}} + \lambda_{4} \mathcal{L}_{\text{struct}} + \lambda_{5} \mathcal{L}_{\text{e}}.
\label{eq:loss}
\end{equation}}
Weighting coefficients $\lambda_i$ are tuned separately for inversion and editing tasks. Definitions of all loss terms and weighted terms are provided in the supplementary material.


\begin{figure*}[t]
   \includegraphics[width=\textwidth]{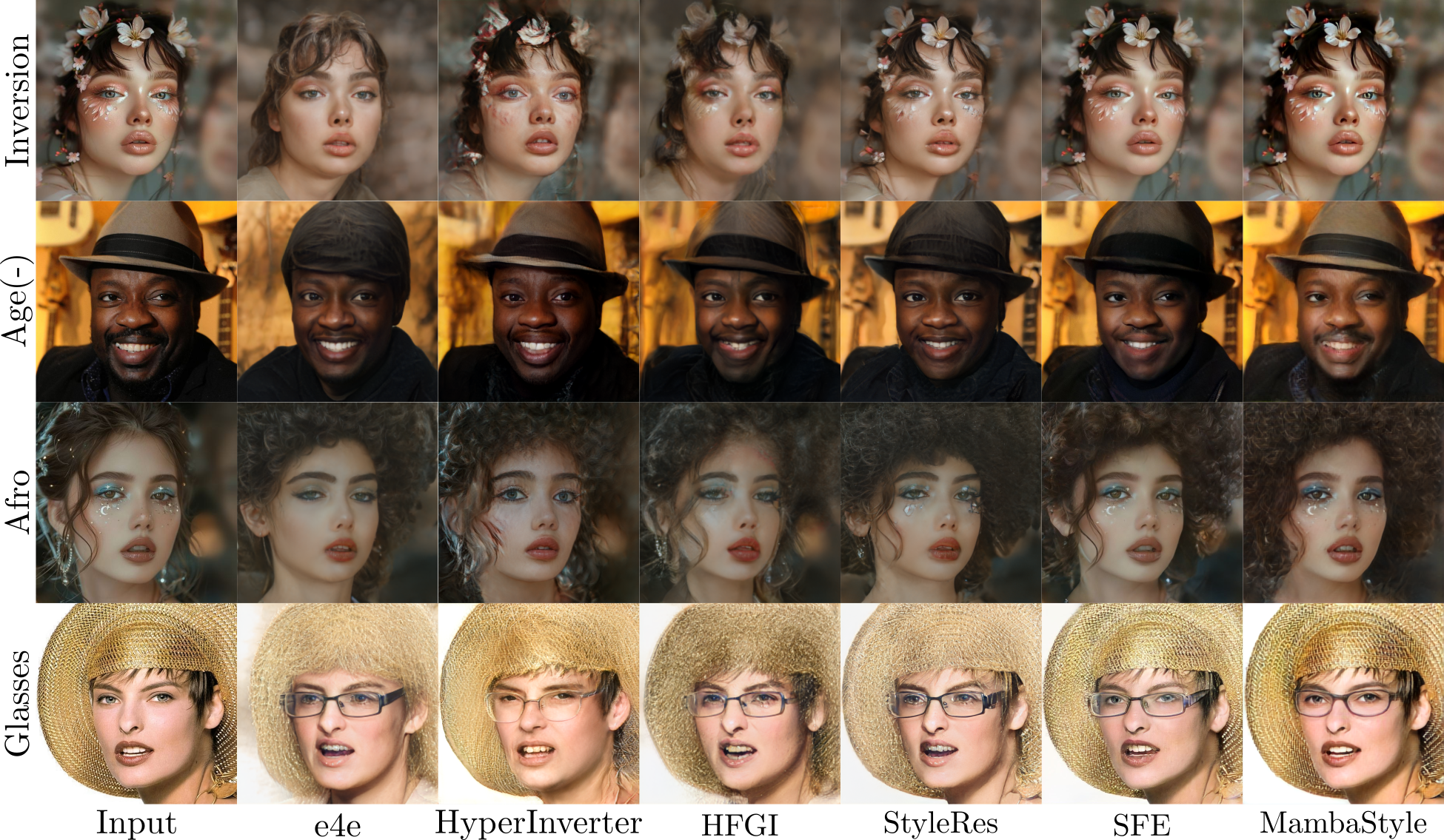}
  \caption{Visual comparison of our proposed method with prior encoder-based approaches in the face domain, based on inversion reconstruction and face editability. Row 1: inversion; Row 2: increasing age; Row 3: Afro hairstyle; Row 4: glasses addition.}
  \vspace{-5pt}
  \label{fig:qualitative_main}
\end{figure*}

\begin{table*}[t]
\centering
\resizebox{\textwidth}{!}{%
\begin{tabular}{ccccccccccccc}
\hline
                                                  & \multicolumn{4}{c}{\textbf{Inversion quality}}                                                                             &                         & \multicolumn{3}{c}{\textbf{Editing quality}}                                                  &                         & \multicolumn{3}{c}{\textbf{Computational Cost}}                                                            \\ \cline{2-5} \cline{7-9} \cline{11-13} 
Model                                             & LPIPS $\downarrow$           & L2 $\downarrow$              & FID $\downarrow$              & MS-SSIM $\uparrow$           &                         & Smile (-)                     & Glasses (+)                   & Old ( + )                     &                         & Params(M)                             & GMACs                                 & Time (s)                   \\ \hline
\rowcolor[HTML]{FFFFFF} 
{\color[HTML]{656565} PTI\cite{roich2022pivotal}} & {\color[HTML]{656565} 0.085} & {\color[HTML]{656565} 0.008} & {\color[HTML]{656565} 14.466} & {\color[HTML]{656565} 0.781} & {\color[HTML]{656565} } & {\color[HTML]{656565} 28.302} & {\color[HTML]{656565} 78.058} & {\color[HTML]{656565} 44.856} & {\color[HTML]{656565} } & {\color[HTML]{656565} 30.37} & {\color[HTML]{656565} 59.70} & {\color[HTML]{656565} 124} \\ \hline
e4e\cite{tov2021designing}                        & 0.199                        & 0.047                        & 28.971                        & 0.625                        &                         & 51.245                        & 119.437                       & 68.463                        &                         & 268.57                                & 125.72                                & 0.034                      \\
pSp\cite{richardson2021encoding}                  & 0.161                        & 0.034                        & 25.163                        & 0.651                        &                         & 46.220                        & 105.740                       & 67.505                        &                         & 297.51                                & 146.75                                & 0.034                      \\
StyleTransformer\cite{hu2022style}                & 0.158                        & 0.034                        & 22.811                        & 0.656                        &                         & 32.936                        & 81.031                        & 67.250                        &                         & \textbf{70.99}                           & {\ul 110.6}                                 & {\ul 0.032}                \\
ReStyle\cite{alaluf2021restyle}                   & 0.130                        & 0.028                        & 20.664                        & 0.669                        &                         & 36.365                        & 87.410                        & 56.025                        &                         & 235.96                                & 221.64                                & 0.138                      \\
Padding Inverter\cite{bai2022high}                & 0.124                        & 0.023                        & 25.753                        & 0.672                        &                         & 42.305                        & 98.719                        & 62.283                        &                         & 308.06                                & 153.05                                & 0.034                      \\ \hline
HyperInverter\cite{dinh2022hyperinverter}         & 0.105                        & 0.024                        & 16.822                        & 0.673                        &                         & 41.201                        & 93.723                        & 65.282                        &                         & 807.70                                & 332.59                                & 0.105                      \\
HyperStyle\cite{alaluf2022hyperstyle}             & 0.098                        & 0.022                        & 20.725                        & 0.700                        &                         & 34.578                        & 86.764                        & 49.267                        &                         & 388.59                                & 258.39                                & 0.275                      \\ \hline
HFGI\cite{wang2022high}                           & 0.117                        & 0.021                        & 15.692                        & 0.721                        &                         & 27.151                        & 77.213                        & 51.489                        &                         & 303.75                                & 250.61                                & 0.072                      \\
Feature-Style\cite{yao2022feature}                & 0.067                        & 0.012                        & 10.861                        & 0.758                        &                         & 26.034                        & 85.686                        & 56.050                        &                         & 144.29                                & \textbf{109.11}                          & 0.038                      \\
StyleRes\cite{pehlivan2023styleres}               & 0.076                        & 0.013                        & 8.505                         & 0.797                        &                         & {\ul 24.465}                        & {\ul 73.089}                  & 43.698                        &                         & 388.82                                & 329.65                                & 0.063                      \\
SFE\cite{bobkov2024devil}                         & \textbf{0.019}               & {\ul 0.002}                  & \textbf{3.535}                & {\ul 0.922}                  &                         & \textbf{24.388}               & 73.098                        & \textbf{41.677}               &                         & 429.33                                & 397.27                                & 0.070                      \\
\rowcolor[HTML]{C4E7D4} 
MambaStyle (Ours)                                 & {\ul 0.025}                  & \textbf{0.001}               & {\ul 7.575}                   & \textbf{0.986}               &                         & 27.149                  & \textbf{72.518}               & {\ul 42.819}                  &                         & {\ul 104.56}                                & 157.82                                & \textbf{0.023}             \\ \hline
\end{tabular}%
}
\caption{Quantitative comparison in the face domain of our proposed method  with SOTA, evaluated across multiple metrics, including inversion and editing quality, and computational cost. The best values are highlighted in bold, and the second-best values are underlined.}
\vspace{-10pt}
\label{tab:SOTA}
\end{table*}

\section{Experimental Results}
\label{sec:experiments}

\mysection{Dataset.} As shown in Fig. ~\ref{fig:proposed} and described in Section ~\ref{sec:method}, during training, we use pretrained StyleGAN2 \cite{karras2020analyzing} models on FFHQ~\cite{karras2019style} and LSUN Cars ~\cite{yu2015lsun} datasets to generate paired data samples, $X$ and $X_e$. These samples are generated from noise vectors $z$ without directly using the original images from the respective dataset. During inference, we follow the same pipeline, but the encoder uses real images as $X$ instead of images generated from noise. Following previous works, we evaluate our model on the CelebA-HQ \cite{karras2017progressive} official face test set and the Stanford Cars \cite{krause20133d} test set for cars. To perform test edits, we use InterfaceGAN \cite{shen2020interpreting} directions for both face and car domains, and StyleClip \cite{patashnik2021styleclip} and GANSpace \cite{harkonen2020ganspace} directions for the face domain only.


\mysection{Metrics.} Following previous works \cite{bobkov2024devil, pehlivan2023styleres, wang2022high}, we assess inversion quaity using LPIPS \cite{zhang2018unreasonable}, L2, and MS-SSIM \cite{wang2003multiscale} metrics, and evaluate the quality of the synthesized images by calculating the Frechet Inception Distance (FID) between the distributions of real and inverted images \cite{heusel2017gans}. Estimating the quality of image edits numerically can be challenging without target images. To address this, we use the approach proposed in \cite{pehlivan2023styleres}. First, we identify the attribute to be edited. Then, using the CelebA-HQ markup, we split the test dataset into two sets: $\mathcal{A}$, containing images with the attribute, and $\mathcal{B}$, containing images without it. We apply our proposed method to the set $\mathcal{B}$ to add the attribute, generating a modified set $\hat{\mathcal{B}}$. The FID score between $\mathcal{B}$ and $\hat{\mathcal{B}}$ is used to evaluate the effectiveness of the method for editing this attribute. We conduct experiments with three attributes: removing smiles, adding glasses, and increasing age. Finally, to assess model efficiency, we measure the total number of parameters required for inference (including StyleGAN parameters), GMACs, and inference time needed to process a single image. GMACs (Giga Multiply-Accumulate Operations) quantify the computational cost by counting the number of operations required to process an image, providing a measure of computational efficiency. 

\mysection{Evaluation Protocol.} We compare our proposed MambaStyle method with several state-of-the-art encoder-based approaches, including e4e \cite{tov2021designing}, pSp \cite{richardson2021encoding}, StyleTransformer \cite{hu2022style}, ReStyle \cite{alaluf2021restyle}, PaddingInverter \cite{bai2022high}, HyperInverter \cite{dinh2022hyperinverter}, Hyperstyle \cite{alaluf2022hyperstyle}, HFGI \cite{wang2022high}, CLCAE \cite{liu2023delving}, Feature-Style \cite{yao2022feature}, StyleRes \cite{pehlivan2023styleres}, and SFE \cite{bobkov2024devil}. Additionally, We also compare our proposed method with the optimization-based method PTI \cite{roich2022pivotal}. We report comparisons on the test sets of the CelebA-HQ and Stanford Cars datasets using the metrics mentioned above. In general, we use the original checkpoints provided by the authors; however, some of these checkpoints are not publicly available for the car domain. Therefore, we exclude models without official checkpoints, and for some methods, such as Feature-Style, we used the values reported in \cite{bobkov2024devil}.


\begin{figure}
 \includegraphics[width=\columnwidth]{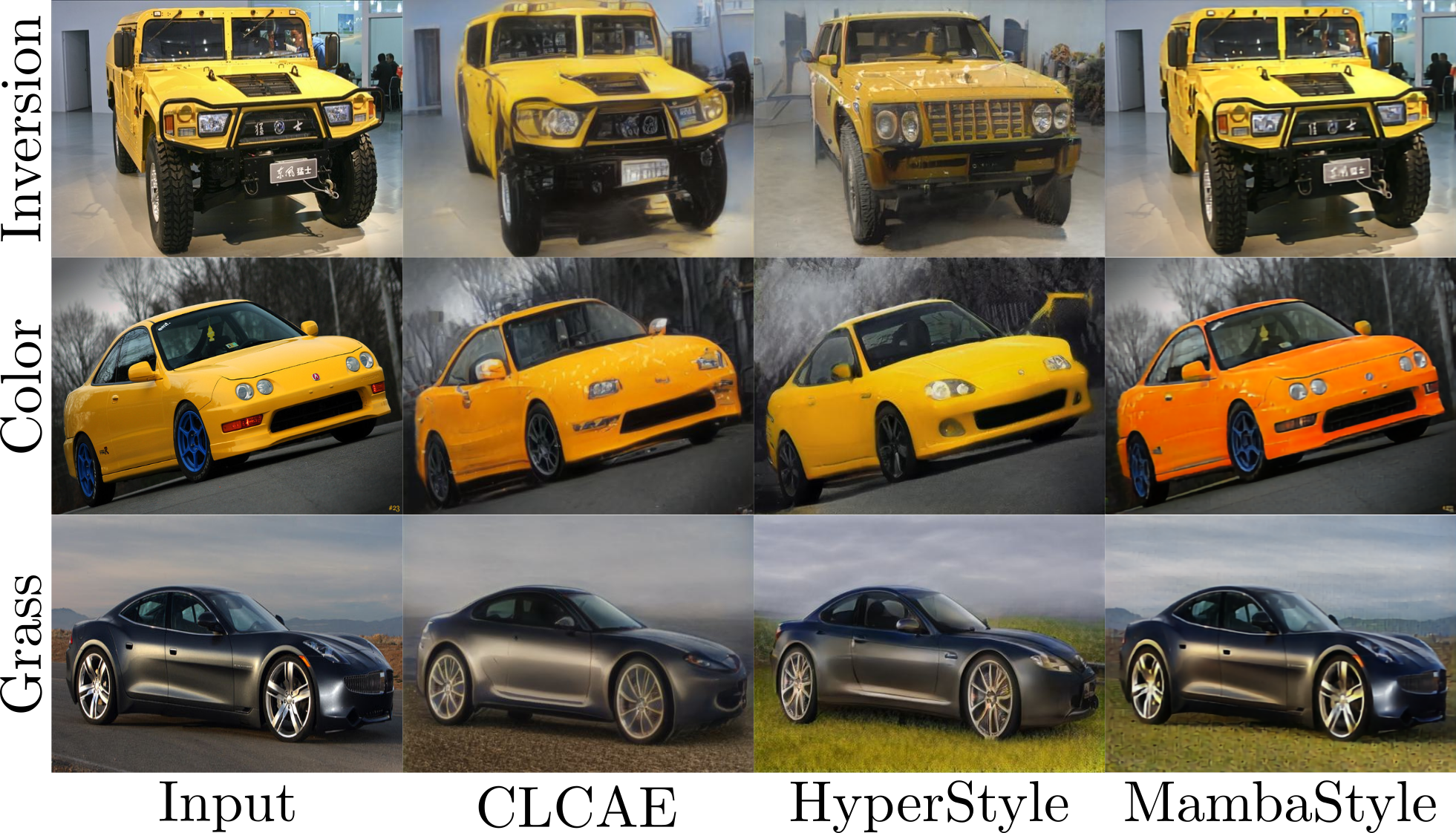}
 \caption{Visual comparison of our proposed method with previous encoder-based approaches in the car domain. Row 1: Inversion; Row 2: Color change; Row 3: grass addition.}
\label{fig:qualitative_add}
\vspace{-15pt}
\end{figure}

\subsection{Qualitative Results}

Figure ~\ref{fig:qualitative_main} visually demonstrates MambaStyle effectiveness in image inversion and editing tasks compared with previous methods. As observed in the first row of Fig. ~\ref{fig:qualitative_main}, MambaStyle accurately reconstructs the \textit{hair flowers} in women, jointly with consistent makeup details. In the second row, MambaStyle shows notable advantages in identity preservation, attribute consistency (e.g., maintaining the appearance of the \textit{hat}), and overall image quality. In contrast, other methods fail to preserve the hat shape, incorrectly reconstructing it with a different color. In the third row, our model produces a more realistic edit of the \textit{afro} hairstyle while preserving original attributes such as hair color, makeup, and eye position. Finally, in the fourth row, all compared methods struggle to generate realistic glasses and significantly fail to preserve the person's identity, including facial details such as the shape and position of the mouth, as well as the fine details of the woven straw hat. Additionally, Fig. ~\ref{fig:qualitative_add} presents a qualitative comparison of our method on the \textit{cars} domain, trained without the $\mathcal{L}_{id}$ loss term (see supplementary for additional results). As observed, our method effectively preserves the background of images both in inversion and after editing. It also reliably maintains key attributes of the input cars, such as size, color, and brightness, during inversion tasks. When applying edits, as color change, our method also preserves attributes better than other approaches, such as HyperStyle, which often fail. 




\subsection{Quantitative Results} \label{rstl:quantitative}
\begin{table}[b!]
\vspace{-10pt}
\centering
\resizebox{0.75\columnwidth}{!}{%
\begin{tabular}{cccc}
\hline
\textbf{Model}   & \textbf{L2}    & \textbf{LPIPS} & \textbf{FID}   \\ \hline
e4e              & 0.122          & 0.325          & 13.397         \\
StyleTransformer & 0.092          & 0.276          & 10.644         \\
ReStyle          & 0.102          & 0.306          & 13.397         \\ \hline
HyperStyle       & 0.080          & 0.287          & 8.044          \\ \hline
Feature-Style    & 0.045          & 0.147          & 7.180          \\
CLCAE            & 0.041          & 0.177          & 11.156         \\
SFE              & {\ul 0.004}    & {\ul 0.039}    & {\ul 4.035}    \\
\rowcolor[HTML]{C4E7D4} 
MambaStyle(ours) & \textbf{0.002} & \textbf{0.031} & \textbf{3.667} \\ \hline
\end{tabular}%
}
\caption{Quantitative comparison of inversion quality between our proposed method and SOTA approaches in the car domain.}
\label{tab:car_results}
\end{table}

\begin{figure*}[t!]
   \vspace{-5pt}
   \includegraphics[width=\textwidth]{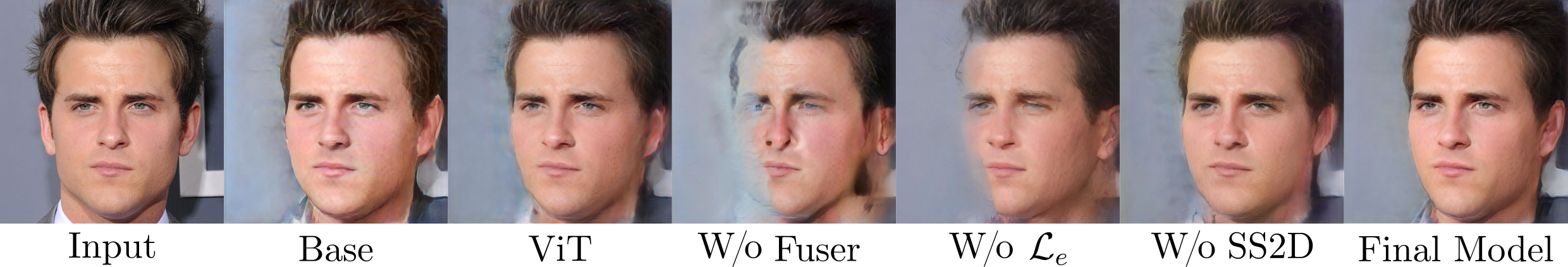}
  \caption{Visual representation of pose rotation across architectural variations of the proposed MambaStyle network.}
  \vspace{-5pt}
  \label{fig:qualitative_ablation}
\end{figure*}

\begin{table*}[t!]
\centering
\resizebox{\linewidth}{!}{%
\begin{tabular}{ccccccccccccc}
\hline
              & \multicolumn{4}{c}{\textbf{Inversion}}                            &           & \multicolumn{3}{c}{\textbf{Editing}}                &           & \multicolumn{3}{c}{\textbf{Computational Cost}}   \\ \cline{2-5} \cline{7-9} \cline{11-13} 
Model         & L2             & LPIPS          & FID            & MS-SSIM        & \textbf{} & Smile(-)        & Glasses(+)      & Old(+)          & \textbf{} & GMacs           & Params(M)      & Time(s)        \\ \hline
Base          & 0.005          & 0.058          & 12.387         & 0.972          &           & 32.628          & 90.950          & 59.230          &           & {\ul 136.33}    & {\ul 89.47}    & \textbf{0.017} \\
ViT           & {\ul 0.003}    & {\ul 0.037}    & {\ul 11.724}   & {\ul 0.978}    &           & 31.674          & 85.943          & {\ul 49.170}    &           & 179.35          & 108.99         & 0.031          \\
W/o Fuser     & 0.013          & 0.072          & 16.368         & 0.935          &           & 40.369          & 83.329          & 60.263          &           & \textbf{110.17} & \textbf{87.12} & {\ul 0.018}    \\
W/o $\mathcal{L}_e$ & 0.005          & {\ul 0.037}    & 14.168         & 0.977          &           & {\ul 28.801}    & {\ul 79.433}    & 55.899          &           & 157.82          & 104.56         & 0.023          \\
W/o SS2D      & 0.004          & 0.046          & 13.584         & 0.973          &           & 43.381          & 85.918          & 57.539          &           & 153.30          & 101.13         & 0.019          \\
\rowcolor[HTML]{C4E7D4} 
Final Model   & \textbf{0.001} & \textbf{0.025} & \textbf{7.575} & \textbf{0.986} & \textbf{} & \textbf{27.149} & \textbf{72.518} & \textbf{42.819} &           & 157.82          & 104.56         & 0.023          \\ \hline
\end{tabular}}%
\caption{Quantitative results from ablation studies in face domain, evaluating the impact of individual components in the proposed architecture. Metrics demonstrate the contribution of each module to the overall MambaStyle performance.}
\vspace{-10pt}
\label{tab:ablations}
\end{table*}

We quantitatively compare our proposed approach to state-of-the-art (SOTA) methods on the official CelebA-HQ test set, using the metrics described in Section ~\ref{sec:experiments} and focusing on three key aspects: inversion quality, editing quality, and computational cost. Table ~\ref{tab:SOTA} summarizes our comprehensive comparison against existing encoder-based SOTA methods, with the results of PTI \cite{roich2022pivotal}, an optimization-based method, included for reference. For inversion quality, our method outperforms SOTA approaches in the L2 and MS-SSIM metrics, which measure pixel-level accuracy and structural similarity, suggesting that our method reconstructs the original image with high accuracy while preserving small details crucial for editing tasks. Although our approach does not outperform SFE in the LPIPS and FID metrics, it achieves the second-best results, with a marginal difference. In terms of editing quality, our method outperforms other approaches in the \textit{Glasses(+)} edit and achieves the second-best results in the \textit{Old(+)} attribute. The reported values correspond to FID scores, as described in Section ~\ref{sec:experiments}. Finally, in terms of computational efficiency, our method significantly outperforms SOTA approaches such as HyperInverter, HyperStyle, StyleRes, and SFE, with a lower number of parameters and GMACs. Although SFE slightly outperforms our method in some specific metrics, our approach offers a notably more efficient solution. Furthermore, the high runtime efficiency provided by the VSSM blocks enables our approach to outperform the SOTA in inference time, making it particularly suitable for real-time applications. Additionally, Table ~\ref{tab:car_results} presents a quantitative comparison of our method with SOTA methods in the car domain, demonstrating that MambaStyle outperforms existing methods across all metrics, establishing a new benchmark in GAN-based inversion and editing. In summary, our approach achieves the best balance between high-quality inversion, editing performance, and model efficiency.



\subsection{Ablation Studies}

We conduct a series of experiments to investigate the contribution of each component of our proposed MambaStyle framework and present qualitative and quantitative results in Fig. ~\ref{fig:qualitative_ablation} and Table ~\ref{tab:ablations}, respectively. In the experiments, the \textit{Base} model refers to our overall framework (shown in Fig. ~\ref{fig:proposed}) with all VSSM and SSM layers removed from both the Encoder and Fuser. In the \textit{ViT} model, we replace the SSM layer in the Fuser with a transformer block and all VSSM modules in the Encoder with vision transformers (ViT) blocks, ensuring the number of parameters remains comparable. This substitution aims to investigate the superior accuracy and lower computational cost of state-space models compared to transformers for the GAN inversion task. Similarly, in the \textit{W/o SS2D} model, we remove the SS2D submodule from the VSSM module to further investigate the importance of 2D selective scanning in our architecture. In the \textit{W/o Fuser} model, the $\hat{H}_3$ feature map is fed directly into StyleGAN, avoiding the Fuser module entirely. This modification aims to demonstrate the importance of our proposed Fuser module, which captures fine details from the original image while integrating the desired edits. In the \textit{W/o $\mathcal{L}_e$} model, we use the complete architecture of our proposed MambaStyle framework but exclude the editing loss $\mathcal{L}_e$ during training to assess its effectiveness in regularizing the trained model. Finally, the \textit{Final Model} represents our full MambaStyle architecture.

As observed in Table ~\ref{tab:ablations}, using \textit{ViT} instead of VSSM in our proposed framework achieves the second-best results quantitatively, with a slightly higher inference time, and significantly outperforms StyleTransformer method \cite{hu2022style}, which relies on a multi-stage transformer-based model. Although the \textit{Base} model achieves the best inference time and second-best results in GMACs and parameter count, it has lower performance across the metrics, indicating that the additional components are required for improving inversion quality and editability. The \textit{W/o Fuser} model indicates that removing the Fuser module significantly reduces computational cost but also leads to a marked drop in performance, demonstrating the Fuser's role in preserving fine details and supporting effective editing. The experiments with the \textit{W/o $\mathcal{L}_e$} model highlight the importance of this loss term, since including $\mathcal{L}_e$ improves the integration of edits and enhances editing quality. Lastly, experiments with the \textit{W/o SS2D} model show that removing SS2D blocks slightly reduces both parameters and GMACs but comes at the cost of performance, highlighting the contribution of SS2D blocks to achieving better feature representation. The visual outputs of the different ablated models, shown in Fig. ~\ref{fig:qualitative_ablation}, highlight that removing the Fuser or the $\mathcal{L}_e$ term significantly reduces editing quality and introduces noticeable artifacts. While the ViT, W/o SS2D, and Base models yield better editing quality, they still produce some artifacts. Overall, our \textit{Final Model} achieves the best balance of inversion quality, editing quality, and computational efficiency. Please refer to the supplementary material to see more ablation experiments.

\begin{figure}[b!]
\vspace{-10pt}
    \centering
    \includegraphics[width=\linewidth]{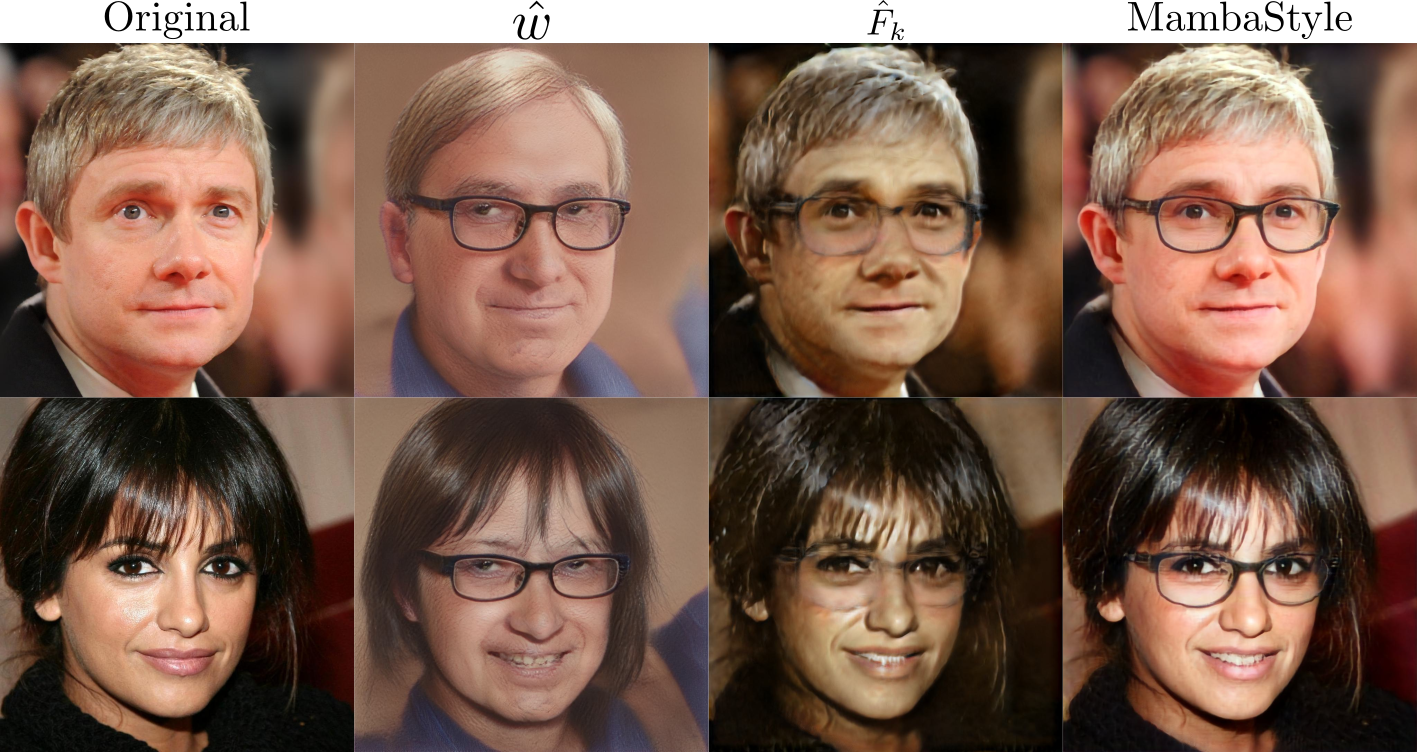}
    \caption{Qualitative results highlighting the importance of the Fuser module in glasses addition tasks.}
    \label{fig:fuser_main}
\end{figure}

\mysection{Fuser Relevance:} A key challenge in GAN-based editing is preserving fine details while applying editions. Our proposed \textit{Fuser} module addresses this by integrating the editing direction $d$ into spatial feature maps, enabling localized feature modulation. Unlike traditional latent-based approaches \cite{tov2021designing, richardson2021encoding, hu2022style}, which apply global transformations, or SFE \cite{bobkov2024devil}, which extracts editing features from a pretrained encoder, our Fuser selectively refines edits while preserving structural fidelity. This design mitigates undesired attribute modifications by restricting editions to relevant regions.

Quantitative results in Tab. \ref{tab:ablations} (row \textit{W/o Fuser}) confirms the Fuser's importance, while Fig.~\ref{fig:fuser_main} visually demonstrates its contribution. The $w'$ column shows results without Fuser, where editing is applied solely through the latent code $w'$, leading to detail loss (e.g., textures, background, skin tone). In contrast, the $\mathcal{F}_k$ column, generated using only the Fuser (with $w'$ set to zero-vector), highlights its ability to retain fine details while producing subtle edits (e.g., slightly blurry glasses). By combining $\hat{w}$ and $\hat{F}_k$ through both the \textit{Encoder} and \textit{Fuser}, our MambaStyle framework preserves details while achieving high-quality edits. Additional analyses are provided in the supplementary material.

\section{Conclusion}
\label{sec:conclusion}


In this paper, we introduce \textbf{MambaStyle}, a novel and efficient single-stage encoder-based approach for GAN inversion and image editing. We leverage vision state-space model modules within an encoder architecture for GAN inversion, enabling efficient learning of latent representations in StyleGAN's latent space and achieving a strong balance between inversion accuracy and editability. Additionally, our approach addresses efficiency by reducing model parameters, inference time, and computational complexity compared to state-of-the-art encoder-based GAN methods. To further enhance detail preservation and mitigate attribute entanglement, we propose the \textbf{Fuser} module, which integrates editing information directly into spatial feature maps, ensuring localized, fine-grained transformations. Our experiments demonstrate that \textbf{MambaStyle} outperforms existing methods in both computational efficiency and the quality of inversion and editing results.

\newpage
\noindent\textbf{Acknowledgments.} This work is supported by the KAUST Center of Excellence for Generative AI under award number 5940. The computational resources are provided by IBEX, which is managed by the Supercomputing Core Laboratory at KAUST.
{
    \small
    \bibliographystyle{ieeenat_fullname}
    \bibliography{main}
}


\end{document}